\documentclass[aps,twocolumn,showpacs,prl]{revtex4}
\usepackage{graphicx,color}
\newlength{\figwidth}
\begin{document}
\title{Finite temperature Mott transition
in Hubbard model on anisotropic triangular lattice}
\author{Takuma Ohashi}\altaffiliation[Present address: ]{Department of Physics, Osaka University, Toyonaka, Osaka 560-0043, Japan. }
\author{Tsutomu Momoi}
\affiliation{%
Condensed Matter Theory Laboratory, RIKEN,
Wako, Saitama 351-0198, Japan}%
\author{Hirokazu Tsunetsugu}
\affiliation{%
Institute for Solid State Physics,
University of Tokyo,
Kashiwa, Chiba 277-8581, Japan}%
\author{Norio Kawakami}
\affiliation{%
Department of Physics, Kyoto University,
Kyoto 606-8502, Japan}%
\date{\today}
\begin{abstract}
We investigate the Hubbard model on the
anisotropic triangular lattice by means of the cellular dynamical 
mean field theory. The phase diagram determined in the Hubbard 
interaction versus temperature plane shows novel reentrant behavior 
in the Mott transition due to the competition between 
Fermi-liquid formation and 
magnetic correlations under geometrical frustration. We demonstrate
 that the reentrant behavior is characteristic of the Mott transition with
intermediate geometrical frustration and indeed consistent with recent experimental 
results of organic materials. 
\end{abstract}
\maketitle

Geometrical frustration has attracted much interest in the field of
strongly correlated electron systems. The discovery of heavy fermion
behavior in the pyrochlore compound $\mathrm{LiV_2O_4}$
\cite{kondo97} and the superconductivity in the triangular lattice
compound $\mathrm{Na_xCoO_2 \cdot yH_2O}$ \cite{takada03} have
stimulated intensive studies of frustrated electron systems.
Geometrical frustration has also uncovered new aspects of the Mott
metal-insulator transition, which is now one of the central issues
in the physics of strongly correlated electron systems. In
particular, recent experiments on the triangular lattice organic
materials $\kappa$-(BEDT-TTF)$_2\mathrm{X}$ have revealed various
interesting physics, such as a novel spin liquid state in the Mott
insulating phase, unconventional superconductivity, etc
\cite{lefebvre00,shimizu03,kagawa04}.
Since the path integral
renormalization group study of the triangular lattice Hubbard model
\cite{kashima01},
the correlated electrons on the anisotropic triangular lattice have
been intensively studied so far
\cite{morita02,imai02,onoda03prb,onoda03jpsj,parcollet04,watanabe04,
watanabe06,yokoyama06,kyung06prl,koretsune07}.
Geometrical frustration effects on the finite-temperature ($T$)
Mott transition, however, have not been sufficiently understood yet.
One of interesting and nontrivial features in the finite-$T$
Mott transition is
reentrant behavior observed in the frustrated
organic material $\kappa$-(BEDT-TTF)$\mathrm{_2Cu[N(CN)_2]Cl}$
under pressure \cite{lefebvre00,kagawa04}.
With lowering temperature,
it once undergoes a transition from Mott
insulator to metal, and then reenters the {\it paramagnetic} insulating
phase at a much lower temperature.
This reentrant behavior is
quite different from the Mott transition in the three dimensional
systems, such as $\mathrm{V_2O_3}$, and is expected to be a new
aspect due to the geometrical frustration and/or low-dimensionality.

The dynamical mean field theory (DMFT) has given 
substantial theoretical progress in understanding the Mott transition,
but this method, not treating spatial fluctuations,
does not explain the reentrant behavior: In DMFT, the metallic phase 
always lies on the lower
temperature side of the finite-$T$ Mott transition line \cite{georges96}.
It is thus desirable to properly incorporate spatially extended 
correlations and geometrical frustration by employing a 
different appropriate method. 

\begin{figure}[bt]
\begin{center}
\end{center}
\includegraphics[clip,width=\figwidth]{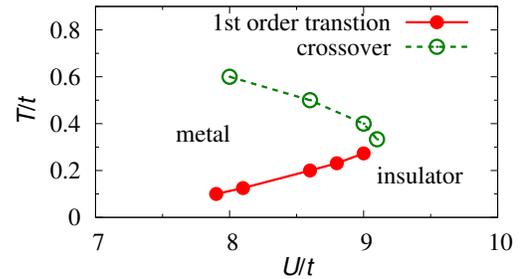}
\caption{
Phase diagram of Hubbard model on
anisotropic triangular lattice
for $t'/t=0.8$.
}
\label{fig:phase}
\end{figure}

In this paper, we study the finite-$T$ Mott transition
in the Hubbard model on the anisotropic triangular lattice
by means of a cluster extension of DMFT,
the cellular-DMFT (CDMFT) \cite{lichtenstein00,kotliar01}. 
The obtained phase diagram
indeed shows a reentrant behavior in the Mott transition
(see Fig. \ref{fig:phase}), 
which is
consistent with that observed
in the organic material mentioned above.  We argue that
the reentrant behavior is a generic feature inherent in
 the Mott transition with intermediate geometrical frustration, and thus
 can be found in various frustrated electron
 systems experimentally.

\begin{figure}[bt]
\begin{center}
\end{center}
\includegraphics[clip,width=\figwidth]{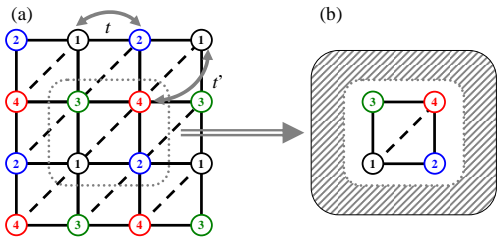}
\caption{
(a) Sketch of the anisotropic triangular lattice and
(b) the effective cluster model using four sites cluster CDMFT.
}
\label{fig:lattice}
\end{figure}

We consider the standard Hubbard model
on the triangular lattice with the hopping anisotropy (see Fig. \ref{fig:lattice}(a)),
\begin{eqnarray}
H= - t  \sum_{\left \langle i,j \right \rangle ,\sigma}
     c_{i\sigma }^\dag c_{j\sigma}
   - t' \sum_{\left( i,j \right) ,\sigma}
     c_{i\sigma }^\dag c_{j\sigma}
   + U \sum_{i} n_{i\uparrow} n_{i\downarrow} ,
\label{eqn:hm}
\end{eqnarray}
with $n_{i\sigma}=c_{i\sigma}^\dag c_{i\sigma}$,
where $c_{i\sigma }^\dag$ ($c_{j\sigma}$) creates (annihilates)
an electron with spin $\sigma$ at site $i$.
To analyze this model we use CDMFT,
which has been successfully applied to frustrated systems such as
the Hubbard model on the triangular lattice \cite{parcollet04,kyung06prl,kyung07} and the
Kagom{\'e} lattice \cite{ohashi06}.
In CDMFT, the original lattice is regarded as a superlattice consisting of clusters,
which is then mapped onto an effective cluster model via a standard DMFT procedure.
Considering four sublattices labeled by 1-4, 
as shown in Fig. \ref{fig:lattice}(a), 
the original lattice model is converted into a four-site cluster model 
coupled to the self-consistently determined medium
illustrated in Fig. \ref{fig:lattice}(b).
Given the Green's function for the effective medium,
$\hat{\mathcal{G}}$,
we can compute the cluster Green's function $\hat{G}$
and the cluster self-energy $\hat{\Sigma}$
by simulating the effective cluster model with
Quantum Monte Carlo (QMC) method \cite{hirsch86}.
Here, $\hat{\mathcal{G}}$, $\hat{G}$,
and $\hat{\Sigma}$ are described by $4 \times 4$ matrices.
The effective medium $\hat{\mathcal{G}}$
is then computed via the Dyson equation,
\begin{eqnarray}
\hat{\mathcal{G}}^{-1} \left( \omega \right) =
\left [ \sum_\mathbf{K}
  \frac{1}{\omega + \mu - \hat{t} \left( \mathbf{K} \right)
  - \hat{\Sigma} \left( \omega \right)}
\right ] ^{-1}
+ \hat{\Sigma} \left( \omega \right) ,
\label{eqn:cavity}
\end{eqnarray}
where $\mu$ is the chemical potential and
$\hat{t} \left( \mathbf{K} \right)$ is the
Fourier-transformed hopping matrix for the sublattice.
Here, summation of $\mathbf{K}$ is taken over the reduced Brillouin zone.
We repeat this procedure until the results are converged
(fifty times iterations are enough).
In each iteration,
we typically use $10^6$ QMC sweeps and
Trotter time slices $L = 12t/T$, where
time slice errors are reduced by interpolation scheme
based on a high-frequency expansion of Green's function \cite{oudovenko02}.

\begin{figure}[bt]
\begin{center}
\end{center}
\includegraphics[clip,width=\figwidth]{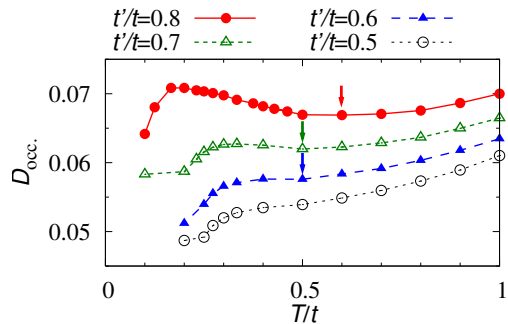}
\caption{ Nonmonotonic temperature dependence of double occupancy
for $U/t=8.0$, where ground states are insulating and in the
vicinity of the Mott transition. 
Arrows denote the crossover temperature 
from the high-$T$ insulator to 
the intermediate-$T$ metal. 
\label{fig:double}} 
\end{figure}

Let us now study the finite-$T$ Mott transition in the model
(\ref{eqn:hm}) at half filling. We first investigate the
$T$-dependence of the double occupancy $D_\mathrm{occ.} = \left
\langle n_{i\uparrow} n_{i\downarrow} \right \rangle$ for
interaction strength $U/t=8$, where the ground state is insulating
and in the vicinity of Mott transition point. A remarkable property
in our frustrated system is nonmonotonic $T$-dependence of $D_\mathrm{occ.}$. 
As shown in Fig.\ \ref{fig:double}, $D_\mathrm{occ.}$
decreases at high-$T$, and then shows an upturn at intermediate-$T$
taking a local minimum at $T/t \sim 0.5$, as $T$
decreases. At much lower $T$, $D_\mathrm{occ.}$ 
starts to decrease again, showing a hump structure. 
This nonmonotonic $T$-dependence of $D_\mathrm{occ.}$ indicates that the
system once changes from insulating to metallic and then
reenters the insulating phase as $T$ is lowered. The
behavior is quite different from that in the infinite dimensional
Hubbard model, in which $D_\mathrm{occ.}(T)$ has a
single minimum at the Fermi-liquid (FL) coherence temperature $T_0$. In
the latter case, the system is insulating at $T>T_0$ and FL at
$T<T_0$ \cite{georges96}. The nonmonotonic behavior in our system is
also different from that in the unfrustrated square lattice Hubbard
model. On the square lattice, the FL coherence is
disturbed by the antiferromagnetic (AF) interaction due to the
perfect nesting, which gives rise to monotonic decrease of
$D_\mathrm{occ.}$
\cite{moukouri01,maier05}.
Note that the hump structures in $D_{occ.}$ become more prominent 
and shift to lower temperatures
as $t'$ increases, although they get smeared for $t'/t=0.5$. 
Therefore, the nonmonotonic behavior is a characteristic feature
caused by geometrical frustration. 

\begin{figure}[bt]
\begin{center}
\end{center}
\includegraphics[clip,width=\figwidth]{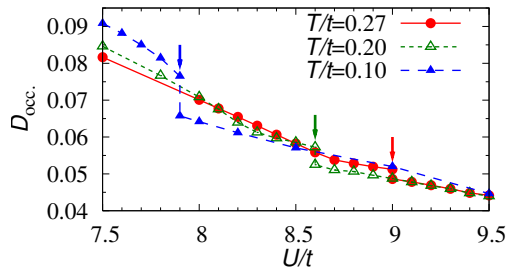}
\caption{ Double occupancy as a function of interaction strength
$U/t$ for $t'/t=0.8$ at several temperatures. 
We show only the transition from insulator to metal with weakening $U$ 
(defined as $U_{c1}$), although we find hysteresis. }
\label{fig:transition}
\end{figure}

To elucidate whether the change between metal and insulator
is a real phase transition or crossover, we further investigate the
double occupancy for typical anisotropy $t'/t=0.8$ with varying $U$. 
We start from the noninteracting system to reach the
large-$U$ regime, typically $U/t=10$, and then calculate
$D_\mathrm{occ.}$ with decreasing $U$ gradually. 
As shown in Fig. \ref{fig:transition}, 
the double occupancy jumps at critical 
interaction strength $U_c$ with decrease of $U$, which signals the
first order Mott transition. 
The jump shrinks with increasing $T$, and
vanishes above $T/t \sim 0.3$. 
The critical end point is expected to be located at  
$T/t \sim 0.3$ and $U/t \sim 9.1$. 
At high-$T$, the system shows
a crossover between metal and insulator, where we define the
boundary by the temperature at which the double occupancy takes the
first local minimum at high-$T$ (see Fig. \ref{fig:double}). 
This boundary is consistent with that determined
by the local minimum of the density of states at the Fermi energy. 
The phase diagram thus determined is drawn in Fig. \ref{fig:phase}. 
The remarkable point is that $U_c(T)$ in our system 
shows a slope with the opposite sign to that in the infinite
dimensional Hubbard model at low temperatures, whereas 
the high-$T$ crossover shows similar behavior to the
well-known results in infinite dimensions.

Within the single-site DMFT, 
entropy per site in the paramagnetic insulating 
phase is roughly $\ln 2$ corresponding to localized free spins 
and the metallic phase with the FL coherence has smaller
entropy. Hence, in the vicinity of Mott
transition temperature, the system becomes insulating at high-$T$
to gain the entropy and the metallic phase always lies
in the lower temperature regime 
\cite{georges96}. However, the spatial fluctuations, which are not
incorporated in the single-site DMFT, become important at low
temperatures. 
For example, the dynamical cluster treatment of
 the unfrustrated square lattice shows that the
FL metallic phase does not appear 
because of strong AF correlations \cite{moukouri01,maier05}. 
On the other hand, in our system, the magnetic correlations are 
hard to develop down to low temperature 
$T/t\sim 0.4$ because of geometrical frustration. Therefore,
as temperature decreases, 
the entropy is released not by spin correlations but
by the itinerancy of electrons in $T/t>0.4$, which causes the crossover 
from insulator to metal in Fig. \ref{fig:phase}.
The appearance of this kind of FL states is one 
of characteristics near the
Mott transition with geometrical frustration \cite{ohashi06}. 
At much lower temperatures, 
the magnetic correlations get enhanced, 
which finally trigger the first order transition from 
the FL to the insulator with smaller entropy. 
Therefore, as $T$ decreases, 
$U_c(T)$ decreases at low temperatures ($T/t<0.4$) 
in contrast to at high temperatures ($T/t<0.4$). 
We thus conclude that the reentrant Mott transition on the anisotropic 
triangular lattice 
is due to the competition induced by geometrical frustration 
between the FL formation and the magnetic correlations.
Since such competition is expected to happen in 
 frustrated electron systems in low dimensions, 
the reentrant behavior in the Mott transition could
be found in a wide variety of frustrated materials.

\begin{figure}[bt]
\begin{center}
\end{center}
\includegraphics[clip,width=\figwidth]{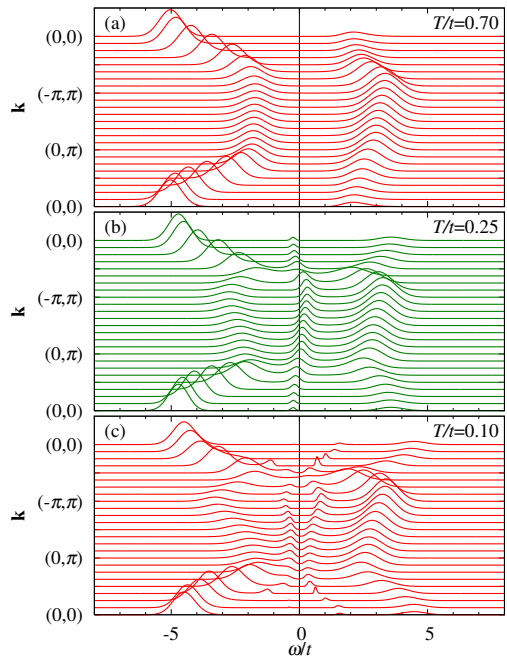}
\caption{
Momentum resolved one-particle spectrum $A_\mathbf{k}(\omega)$
for $U/t=8.0$, $t'/t=0.8$ at several temperatures.
}
\label{fig:arpes}
\end{figure}

We can clearly see the development of the quasiparticles and 
magnetic correlations
discussed above in the momentum resolved one-particle spectrum
$A_\mathbf{k}(\omega)$.
Within CDMFT, the one-particle Green's function is given as,
\begin{eqnarray}
G_\mathbf{k}\left( \omega \right) = \frac{1}{4}
\sum_{\gamma,\delta}
e^{i\mathbf{k} \cdot \left( \mathbf{r}_\gamma - \mathbf{r}_\delta \right)}
\left[ \omega + \mu - \hat{t} \left( \mathbf{k} \right)
  - \hat{\Sigma} \left( \omega \right)
\right ]^{-1}_{\gamma\delta},
\label{eqn:green}
\end{eqnarray}
where $\mathbf{k}$ is the wave vector in the original Brillouin zone
and $\mathbf{r}_\gamma$, $\mathbf{r}_\delta$ label four cluster
sites \cite{kyung06prb}. We calculate the imaginary time Green's
function $G_\mathbf{k}\left( \tau \right)$ and obtain the spectrum
$A_\mathbf{k}\left( \omega \right) = -\mathrm{Im} G_\mathbf{k}\left(
\omega + i0 \right)/\pi$ using the maximum entropy method \cite{jarrell96}.
In Fig. \ref{fig:arpes}, we show
$A_\mathbf{k}\left( \omega \right)$ for $U/t=8$, $t'/t=0.8$
at typical temperatures.
At high temperatures $A_\mathbf{k}\left( \omega \right)$
shows an insulating behavior, where it has
a large Hubbard gap of order of $U/t$ and no quasiparticle peak.
As temperature is lowered, the quasiparticle peak starts to develop 
inside the gap, having weak dispersion,
which clearly demonstrates the emergence of the frustration-induced 
metallic phase.
The stabilization of the metallic phase by geometrical frustration is consistent with
the previous studies of the Hubbard model on the triangular lattice \cite{imai02,maier05}
and the Kagom{\'e} lattice \cite{ohashi06}.
At much lower temperatures, the quasiparticle peaks split and 
acquire a very small gap, where the system is insulating again. 
This small gap is due to the magnetic exchange interaction
among the quasiparticles, which is consistent with the results at zero temperature
recently obtained by CDMFT with exact 
diagonalization method \cite{kyung06prl}. 
We thus confirm that our frustrated system 
exhibits the insulator-metal-insulator reentrant behavior
as temperature decreases.

\begin{figure}[bt]
\begin{center}
\end{center}
\includegraphics[clip,width=\figwidth]{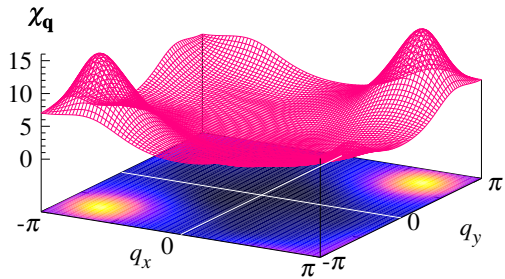}
\caption{
Static spin susceptibility $\chi_q$ in the insulating phase
for $U/t=9$, $t'/t=0.8$ at $T/t=0.2$.
}
\label{fig:chi}
\end{figure}

Finally we examine the magnetic instability by calculating 
the static spin susceptibility $\chi_q$. 
We find that the spin susceptibility $\chi_q$ remains finite just below 
the first order Mott transition temperature and hence
the {\it paramagnetic} Mott insulator is not precluded by the magnetically ordered phase 
in contrast to the single-site DMFT where the magnetic order
almost conceals the Mott transition \cite{georges96,zitzler04}. 
Although the appearance of a finite-temperature magnetic transition is 
due to a mean-field type approximation, 
the magnetic ordering in our case, highly suppressed by frustration effects, 
appears below the Mott transition temperature.
In Fig. \ref{fig:chi}, we show
$\chi_q$ for $U/t=9$, $t'/t=0.8$ at $T/t=0.2$, where the system is
in the insulating phase close to the Mott transition point. The
susceptibility $\chi_q$ takes a maximum at incommensurate wave vectors
$\mathbf{q} \sim (0.7 \pi, 0.7 \pi)$ and does not diverge.
The magnetic correlations thus play an important
role in driving the Mott transition at low temperatures, 
but do not induce a real magnetic instability 
due to strong frustration. 
We note that our calculations of $\chi_q$ for 
different values of $t'$ are also consistent with recent results of
the Hubbard model around the Mott transition \cite{kashima01,morita02,watanabe06,yokoyama06}: 
For large-$U$ and weak frustration $t'/t \le 0.7$, 
$\chi_q$ has a peak at $\mathbf{q} = (\pi, \pi)$ corresponding to the commensurate
antiferromagnetism, which rapidly develops with lowering $T$. 
We also find that $\chi_q$ for $t'/t = 1$ takes a maximum at 
$\mathbf{q} = (2 \pi/3, 2 \pi/3)$ corresponding to the $120^\circ$ structure, 
in which the development of $\chi_q$ is much slower than the AF correlations for small $t'$. 

In summary,
we have investigated the Mott transition in the Hubbard model
on the anisotropic triangular lattice by means of CDMFT.
The obtained phase diagram (Fig. \ref{fig:phase}) in the $U$-$T$ plane for $t'/t=0.8$ 
is qualitatively consistent with the reentrant Mott transition in
$\kappa$-(BEDT-TTF)$\mathrm{_2Cu[N(CN)_2]Cl}$ with $t'/t \sim 0.75$.
By calculating the wave-vector dependent spectral function and 
susceptibility, we have clarified that 
the reentrant-type Mott transition is due to the competition 
between the FL formation and the magnetic correlations
under geometrical frustration. 
We have also studied the Mott transition for different values of $t'/t$. 
For weakly frustrated case $t'/t \le 0.5$, we have found that 
the FL formation is suppressed by the strong AF correlations, so that
the reentrant Mott transition becomes obscured. 
On the other hand, we have found that for fully frustrated case $t' \sim t$, 
the FL states are well stabilized by frustration and 
the metallic region is extended, 
because the magnetic fluctuations of 
the $120^\circ$ structure are weak. 
Hence, the low-$T$ Mott transition line shifts to much lower-$T$ regime 
\cite{parcollet04}. 
The above tendency is consistent with another organic material 
$\kappa$-(BEDT-TTF)$_2\mathrm{Cu_2(CN)_3}$ with $t' \sim t$, 
where magnetic ordering was not observed down to the 
lowest temperature studied experimentally.
Therefore the reentrant behavior can be most clearly 
observed in moderately frustrated electron systems.
Although we have used the four-site-cluster CDMFT, 
we believe that the reentrant behavior of Mott transition is robust 
and our results do not qualitatively change even for a larger cluster size, 
because the competition between FL formation and 
magnetic correlations should occur 
generally in frustrated electron systems. 
We expect that the reentrant behavior in the Mott transition
found here will be observed experimentally in a variety of frustrated 
electron systems.

The authors thank T. Koretsune and K. Inaba for valuable discussions.
A part of numerical computations was done at 
the Supercomputer Center at ISSP, University of Tokyo
and also at YITP, Kyoto University. 
This work was partly supported by 
KAKENHI (17071011, 18043017, and 19052003)
and also by the Next Generation Super Computing Project, 
Nanoscience Program, from the MEXT of Japan.

\end{document}